\documentclass[apj]{emulateapj}
\usepackage{graphicx,subfigure}

\newcommand{\hz}{high-$z$}
\newcommand{\ztwo}{$z \sim 2$}

\newcommand{\ha}{H$\alpha$}				    
\newcommand{\hb}{H$\beta$}					    
\newcommand{\nii}{[N{\small II}]}                                       
\newcommand{\sii}{[S{\small II}]}                                       

\newcommand{\msun}{M$_{\odot}$} 			   
\newcommand{\mstar}{M$_{*}$} 				    
\newcommand{\mbh}{M$_{\rm BH}$} 			    
\newcommand{\mbulge}{M$_{\rm bulge}$} 		    
\newcommand{\mdotbh}{$\dot{\rm M}_{\rm BH}$} 	    

\newcommand{\msunyr}{\msun\ yr$^{-1}$}		    
\newcommand{\kms}{km~s$^{-1}$}			              

\newcommand{\nsins}{47}
\newcommand{\nmass}{45}

\shorttitle{Broad Emission in High-Redshift Galaxies}
\shortauthors{K. L. Shapiro et al.}

\begin{document}


\title{The SINS Survey: Broad Emission Lines in High-Redshift Star-Forming Galaxies\footnotemark[*]}
 
 \footnotetext[*]{Based on observations obtained at the Very Large Telescope (VLT) of the European Southern Observatory, Paranal, Chile in the context of ESO programs 070.A-0229, 070.B-0545, 073.B-9018, 074.A-9011, 075.A-0466, 076.A-0527, 077.A-0576, 078.A-0055, 078.A-0600, 079.A-0341, 080.A-0330, and 080.A-0635.}

\author{Kristen L. Shapiro\altaffilmark{1}, Reinhard Genzel\altaffilmark{2,3}, Eliot Quataert\altaffilmark{1}, 
		Natascha M. F\"orster Schreiber\altaffilmark{3}, Richard Davies\altaffilmark{3},
		Linda Tacconi\altaffilmark{3}, Lee Armus\altaffilmark{4}, Nicolas Bouch\'e\altaffilmark{3},
		Peter Buschkamp\altaffilmark{3}, Andrea Cimatti\altaffilmark{5},
		Giovanni Cresci\altaffilmark{3}, Emanuele Daddi\altaffilmark{6},
		Frank Eisenhauer\altaffilmark{3}, Dawn K. Erb\altaffilmark{7},
		Shy Genel\altaffilmark{3}, Erin K. S. Hicks\altaffilmark{3}, Simon J. Lilly\altaffilmark{8},
		Dieter Lutz\altaffilmark{3},
		Alvio Renzini\altaffilmark{9}, Alice Shapley\altaffilmark{10}, Charles C. Steidel\altaffilmark{11},
		Amiel Sternberg\altaffilmark{12}}
\altaffiltext{1}{Department of Astronomy, University of California, Berkeley, California 94720, USA}
\altaffiltext{2}{Department of Physics, University of California, Berkeley, California 94720, USA}
\altaffiltext{3}{Max-Planck-Institut f\"ur extraterrestrische Physik (MPE), Giessenbachstr.1, D-85748 Garching, Germany} 
\altaffiltext{4}{Spitzer Science Center, California Institute of Technology, Pasadena, CA 91125, USA}
\altaffiltext{5}{Istituto Nazionale di Astrofisica-Osservatorio Astronomico di Bologna, Via Gobetti 101, I-40129 Bologna, Italy}
\altaffiltext{6}{Service d'Astrophysique, Dapnia CEA, Saclay, France}
\altaffiltext{7}{Harvard-Smithsonian Center for Astrophysics, 60 Garden Street, Cambridge, MA 02138 USA}
\altaffiltext{8}{Institute of Astronomy, Eidgen\"ossische Technische Hochschule, ETH Zurich, CH-8093, Switzerland}

\altaffiltext{9}{Osservatorio Astronomico di Padova, Vicolo dell'Osservatorio 5, Padova, I-35122, Italy}
\altaffiltext{10}{Department of Physics and Astronomy, University of California, Los Angeles, CA 90095 USA}
\altaffiltext{11}{California Institute of Technology, MS 105-24, Pasadena, CA 91125, USA}
\altaffiltext{12}{School of Physics and Astronomy, Tel Aviv University, Tel Aviv 69978, Israel}


\begin{abstract}
High signal-to-noise, representative spectra of star-forming galaxies at \ztwo, obtained via stacking, reveal a high-velocity component underneath the narrow \ha\ and \nii\ emission lines.  When modeled as a single Gaussian, this broad component has FWHM~$\gtrsim$~1500~\kms; when modeled as broad wings on the \ha\ and \nii\ features, it has FWHM~$\gtrsim$~500~\kms.  This feature is preferentially found in the more massive and more rapidly star-forming systems, which also tend to be older and larger galaxies.   We interpret this emission as evidence of either powerful starburst-driven galactic winds or active supermassive black holes.  If galactic winds are responsible for the broad emission, the observed luminosity and velocity of this gas imply mass outflow rates comparable to the star formation rate.  On the other hand, if the broad line regions of active black holes account for the broad feature, the corresponding black holes masses are estimated to be an order of magnitude lower than those predicted by local scaling relations, suggesting a delayed assembly of supermassive black holes with respect to their host bulges.
\end{abstract}




\section{Introduction}

Observations of galaxies in the early Universe are a unique probe of matter assembly during its most active epoch; at \ztwo, both the cosmic star formation rate and the luminous quasar space density are at their peaks \citep[e.g.][]{Fan+01,Cha+05}.  Galaxies themselves undergo corresponding growth during this time, with the total stellar mass density in galaxies increasing from $\sim$~15\% to 50$-$75\% its current value between $z \sim$~3 and $z \sim$~1 \citep[e.g.][]{Dic+03,Rud+03,Rud+06}.  Constraining the dynamical and baryonic processes driving this rapid evolution is therefore central to our understanding of galaxy formation and to informing cosmological simulations.

At the relevant epochs, key spectral diagnostic features are redshifted into the near-infrared.  In recent years, a number of surveys have therefore begun to systematically probe high-redshift populations with near-infrared spectroscopy \citep[e.g.][]{Erb+06a,Erb+06c,Erb+06b,Swi+04,Tak+06,Kri+08}.  Our recently completed SINS (Spectroscopic Imaging in the Near-infrared with SINFONI) survey has combined the resolving power of 8-10m class telescopes with high-resolution integral field spectrographs to study the detailed internal processes at work within massive, star-forming galaxies at \ztwo\ (\citealt{For+06,For+09,Gen+06,Gen+08,Bou+07,Sha+08,Cre+09}; see also related work by \citealt{Pue+06,Swi+06,Wri+07,Law+07,Law+09}).

In this paper, we combine the spectra of \nsins\ galaxies detected in \ha\ emission to study the average spectral properties of star-forming galaxies at \ztwo\ (\S\ref{Data}).  In particular, we report the discovery of broad emission lines in these galaxies (\S\ref{Results}) and interpret this high-velocity warm gas as arising either in large-scale galactic winds driven by the high star formation rates (SFR) in these galaxies or in the broad-line regions (BLR) surrounding active galactic nuclei (AGN).  We explore the implications of both scenarios in \S\ref{Discussion} and conclude in \S\ref{Conclu}.

Throughout this paper, we assume a $\Lambda$-dominated cosmology with $H_{\rm 0} =$ 70 km s$^{-1}$ Mpc$^{-1}$, $\Omega_{\rm m} =$~0.3, and $\Omega_{\rm \Lambda} =$~0.7.  For this cosmology, 1$\arcsec$ corresponds to $\approx$~8.2~kpc at $z =$~2.2.

\section{Data and Analysis}
\label{Data}

In the context of the SINS program, 80 $z = $1$-$3 systems were observed in emission lines in the infrared (rest-frame optical) with VLT/SINFONI \citep{Eis+03,Bon+04} for an average of 3.5~hours per band and pixel scale on each target \citep{For+09}.  These galaxies were largely (62/80) taken from the (rest-frame) UV-selected samples of \citet{Erb+06c,Erb+06b} and the (rest-frame) optically-selected samples of \citet{Abr+04}, \citet{Dad+04a}, \citet{Kon+06}, \citet{Lil+07}, and Kurk et al. (in prep).  From these magnitude- and color-defined samples, suitable SINS targets were culled, with the main selection criteria being a combination of target visibility during the observing runs, night sky line avoidance in the emission lines of interest, and an estimated integrated emission line flux $\gtrsim$~5~$\times$~10$^{-17}$ erg s$^{-1}$ cm$^{-2}$, such that high quality data could be obtained in reasonable integration times.  Of the 62 rest-frame UV/optically-selected galaxies chosen in this manner, 52 were well detected in \ha\ in our SINFONI observations.

\citet{For+09} discuss the selection of this sample in detail and show that the SINS galaxies are representative of the \ztwo\ star-forming galaxy population, with some bias towards the more rapidly star-forming (and therefore more luminous in \ha\ emission) systems.  They also note that the SINS sample (and some of the parent samples) selects against known AGN and quasars, in the interest of studying the dynamic and evolutionary state of \ztwo\ star-forming galaxies.  However, a small number of previously known AGN were in fact observed in the SINS program; in the UV/optically-selected part of the sample, there are 5 such systems, as originally identified with the UV/optical spectroscopy of the parent surveys.  For further details about the SINS sample, observations, and data reduction, we refer interested readers to \citet{For+09}.

Here, we analyze our SINS observations of the galaxies that were UV/optically-identified and that are well-detected in \ha\ in individual spatial elements; this population (totalling \nsins\ of 52 sources detected in \ha, including 4/5 of the previously known AGN) comprises the majority of the SINS sample.  To study the average properties of these \ztwo\ star-forming galaxies, we generated stacked, high-signal-to-noise (S/N) spectra representative of the population as a whole.

From our integral field data, we first created a spatially-integrated one-dimensional spectrum for each galaxy by shifting each spectrum within the galaxy datacube by its measured (\ha) velocity and then collapsing the datacube into a single spatially-integrated spectrum.  In this manner, the spatially-integrated spectrum contains no systematic velocity broadening (e.g. by large-scale rotation) on scales larger than the PSF (FWHM~$\sim$~4~kpc).  Testing confirmed that this approach did not affect the properties of the broad \ha\ component (see below) but did improve the signal-to-noise (S/N) of the detection.  This technique has the additional benefit of randomizing OH atmospheric emission lines in the \ha\ rest-frame and therefore effectively eliminating residuals from the OH line removal.  The remaining residuals from this process were inspected and masked out by hand.

The \nsins\ spatially-integrated spectra were then combined into a single spectrum (with equivalent integration time of 195 hours) by interpolating all spectra onto a common wavelength axis, converting their measured fluxes to luminosities using their luminosity distances, weighting each spectrum by the S/N of the \ha\ emission line, and averaging the resulting spectra.  During this process, we do not correct for extinction (but see \S\ref{Discussion}).  Typical extinctions our sample galaxies have been measured to be $A_V \sim$~1 (\citealt{For+09}; see also e.g. \citealt{Dad+04b,Erb+06b}), which translates to an underestimation of our \ha\ luminosities by at most a factor of $\sim$~2.

\begin{figure*}
	\centering
	\includegraphics[width=17cm,trim=0cm 37mm 0cm 0cm]{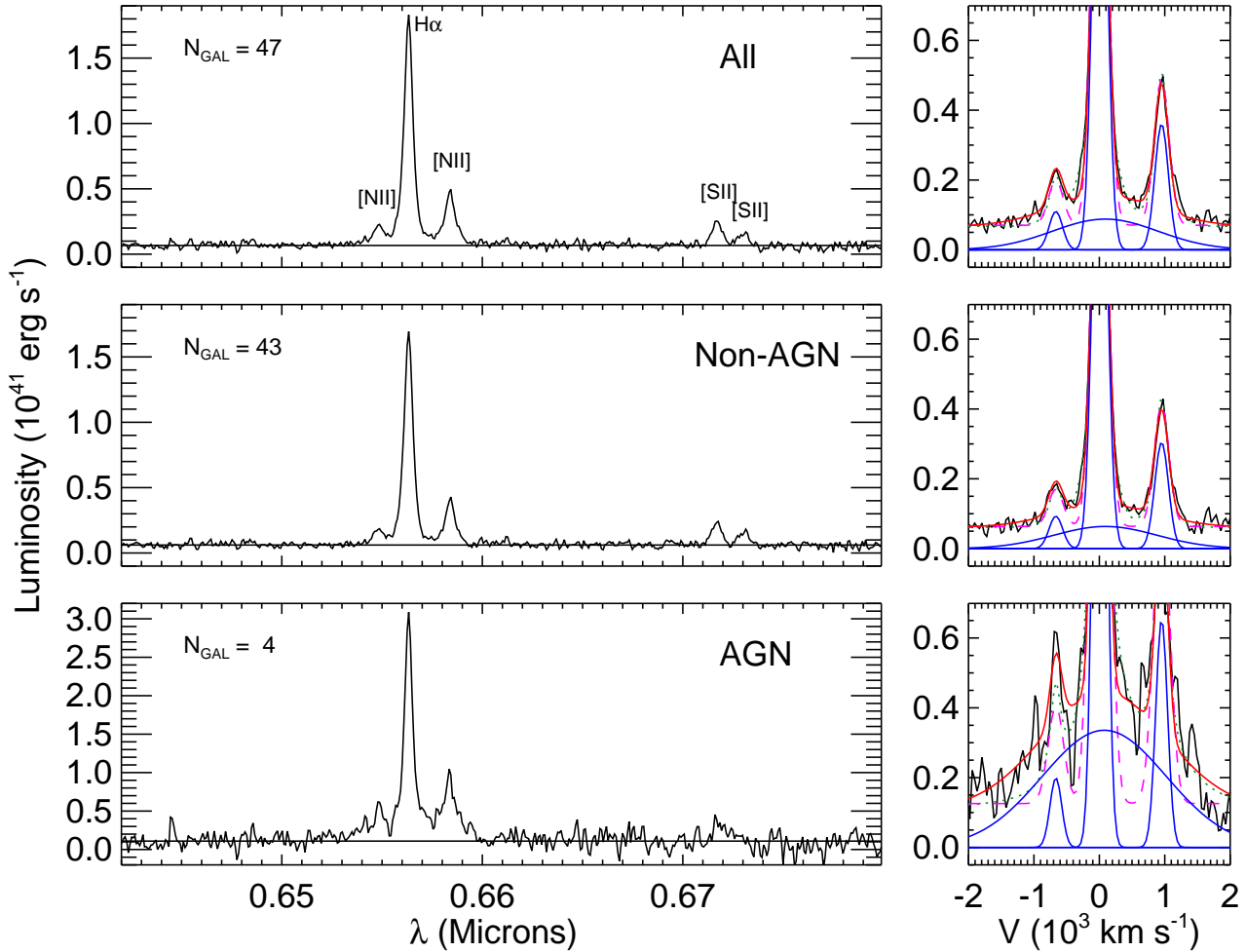}
	\caption{{\it Left}: Average spectrum of the \nsins\ SINS galaxies ({\it top panel}), followed by the average spectrum of all non-active systems and that of all systems previously known to host AGN ({\it lower two panels}).  {\it Right}: Zoomed view of the \ha\ and [N{\tiny II}] region, on a velocity scale, with the best-fitting combination of a constant continuum, narrow lines, and a broad component overplotted ({\it red}).  The individual line components are also plotted ({\it blue}).  For reference, also overplotted are the best fits derived by fitting both \ha\ and [N{\tiny II}] with only narrow lines ({\it dashed magenta}) and with a narrow and a broad component ({\it dotted green}).}
	\label{agn}
\end{figure*}

The average spectrum for the SINS \ztwo\ star-forming galaxies is presented in the top panel of Figure~\ref{agn}.  We also created average spectra of subsets of the SINS galaxy sample, in order to test the dependence of spectral properties on other known galaxy properties.

Our average spectra reveal a broad emission component underneath the bright narrow lines.  We quantify this feature in each average spectrum by simultaneously fitting a combination of a constant continuum offset, narrow lines (\ha, \nii, \sii) of identical kinematics (velocity and velocity dispersion), and a single broad component, whose kinematics are allowed to vary.  All lines are assumed to be well-described by a single Gaussian; the validity of this assumption is confirmed by a reduced $\chi^2$ of close to unity ($\chi_{dof}^2 \sim$ 0.9$-$1.9) for all fits (see Table~\ref{tab:results}).  Our data can also be well fit by fitting a combination of a constant continuum offset, narrow lines with shared kinematics, and broad forbidden and permitted lines with shared kinematics, with the \nii/\ha\ ratio identical in the narrow and broad component.  All lines are assumed to be well-described by this combination of two Gaussians (dotted green line in Figure~\ref{agn}; $\chi_{dof}^2 =$ 1.4), which has the same number of free parameters as the previous fit.  In the limit of the S/N of our data, we cannot add additional free parameters to the fits, nor can we identify a preferred model.  For clarity, we refer throughout to the former (single broad Gaussian under the \ha+\nii\ complex) as ``broad lines" and the latter (double Gaussians for both \ha\ and \nii) as ``broad wings."  For simplicity and for comparison with the literature, we primarily quantify the observed high-velocity feature with a single broad \ha\ line in \S\ref{Results}.  However, we also discuss the implications of the broad wings scenario on the derived properties of the broad emission (\S\ref{Results}) and on the interpretation of this emission (\S\ref{Discussion}).

From these fits to the average spectra, we measure the fractional contribution of broad emission to the overall emission line flux, the kinematics of the broad component, and the line ratios of the narrow line components.  The significance of these measurements are quantified in two ways.  First, for each average spectrum, we re-created 100 spectra by randomly sampling (with replacement) and combining the individual contributing galaxy spectra.  The properties of the broad component were measured in each of the resulting 100 average spectra, yielding confidence intervals for all derived quantities.  Second, the probability of false positive detections $P_{false}$ was tested by creating 1000 simulated spectra with the narrow line and observational properties (noise, spectral resolution) characteristic of each average spectrum.  Comparing the derived broad components in these spectra to those in the real SINS spectra, we estimate the rate of spurious detections of broad components equal to or more prominent (in luminosity and FWHM) than those in the actual data.  The results of this analysis for the different average spectra are presented in the following section.

\section{Results}
\label{Results}

We find that the average SINS galaxy spectrum includes a significant amount of broad emission (top panel of Figure~\ref{agn}), with $\chi^2_{dof}$ = 4.9 for a fit with only narrow lines (dashed magenta line in top right panel of Figure~\ref{agn}) and $\chi^2_{dof}$ = 1.8 and 1.4 for fits including a broad \ha\ line and broad \ha\ and \nii\ wings, respectively (red and dotted green lines in same panel; $P_{false} =$~2\% and 1\%).  To ensure that this signature is not the result of the 4 known AGN included in this sample, we also create stacked spectra of the AGN and the rest of the sample.  While the broad emission from the AGN host systems alone is quite substantial, a comparison of the average SINS spectrum and the average non-AGN spectrum illustrates that the broad emission in the average SINS spectrum is not dominated by that coming from the 4~AGN.  These results are summarized in Table~\ref{tab:results}.

\begin{figure*}
	\centering
	\includegraphics[width=17cm,trim=0cm 37mm 0cm 0cm]{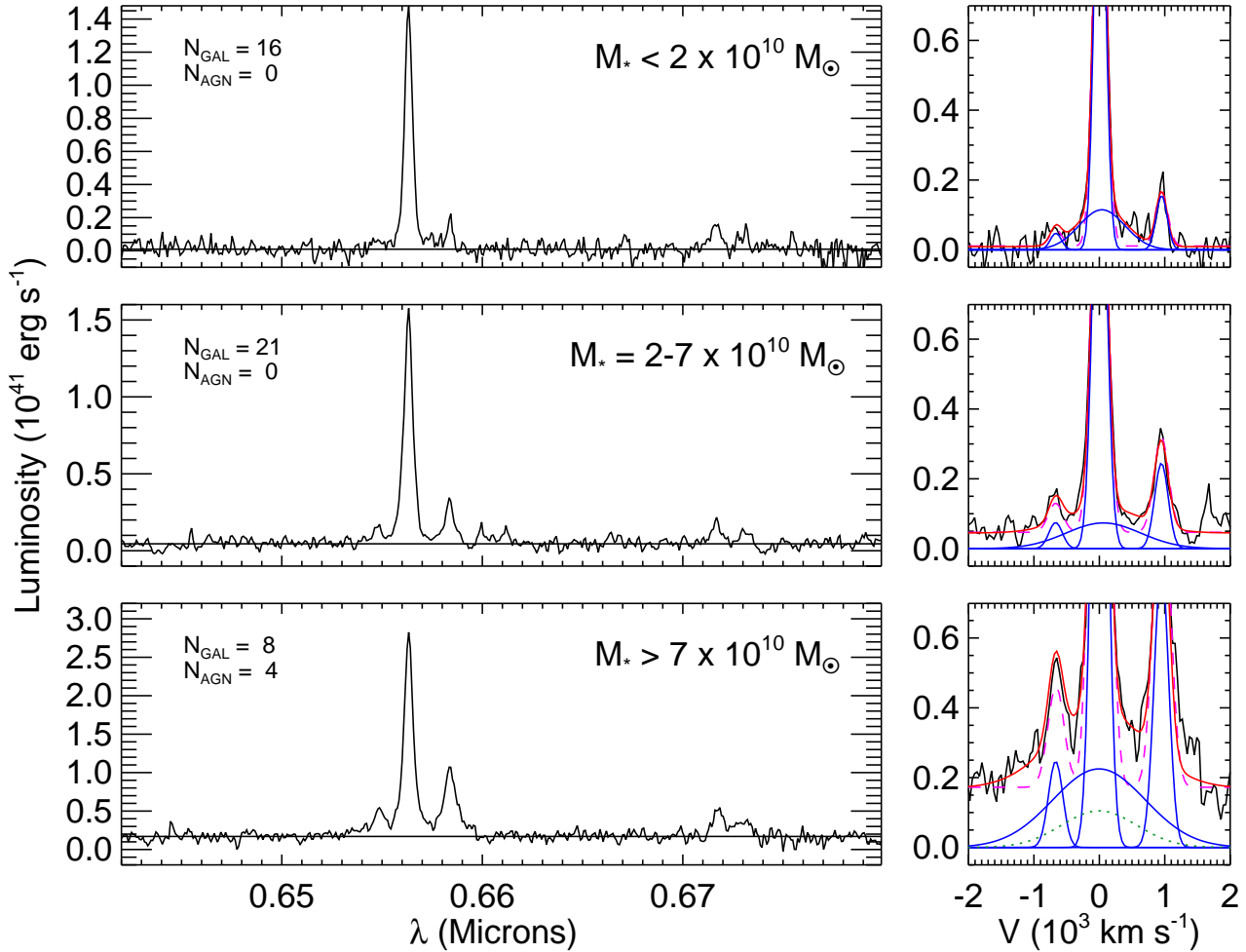}
	\caption{{\it Left}: Average spectrum of each mass bin.  {\it Right:} Zoomed view of the \ha\ and [N{\tiny II}] region, on a velocity scale, with the best-fitting combination of a constant continuum, narrow lines, and a broad component overplotted ({\it red}).  The individual line components are also plotted ({\it blue}).  The best fit obtained with only narrow emission lines is plotted for comparison ({\it dashed magenta}).  In the high mass bin, the best-fit broad line component to the average spectrum of the $3$ non-active high mass galaxies is also overplotted ({\it dotted green}).}
	\label{mstar}
\end{figure*}

We test the dependence of the presence of broad emission on galaxy properties by dividing the sample into three stellar mass bins, using the results from the spectral energy distribution (SED) fitting of \citet{For+09} for the \nmass\ of our \nsins\ targets for which sufficient broadband data exist.  The average spectra for these three bins show an increasing presence of a broad component with stellar mass (Figure~\ref{mstar}; $P_{false} =$~11\%,~5\%,~3\% respectively).  However, the spectrum of the highest mass bin is significantly affected by the contribution of the 4 previously known AGN in our sample, which all fall into this bin.  The nuclear emission in these systems can bias the results of SED fitting towards larger masses, so we confirm their high masses with the dynamical mass measurements made by \citet{Cre+09}; in all cases, the dynamical masses of these galaxies are consistent with the stellar masses used here and remain among the highest in the SINS sample.  Nevertheless, we confirm that a (weaker) broad component is also present in the non-active galaxies in this bin (green line in Figure~\ref{mstar}), with $P_{false} =$~9\%.

Several other key properties of galaxies, including SFR, size, stellar age, and metallicity, have well established correlations with stellar mass in \hz\ galaxies \citep[e.g.][]{Noe+07,Tru+06,Erb+06c}.  Both the SFR-\mstar\ relation and the mass-metallicity relation are apparent in Figure~\ref{mstar}, via the increasing narrow \ha\ luminosity and increasing \nii/\ha\ ratio with stellar mass, respectively.  Since the \nii/\ha\ ratio remains well below levels expected of shock heating or AGN activity, this latter is most likely tracing variations in metallicity (see below, as well as Buschkamp et al.~in prep).  To these established mass-dependent properties in \hz\ galaxies, we now add the presence and strength of a broad component.

\begin{figure*}
	\centering
	\includegraphics[width=10cm, angle=90]{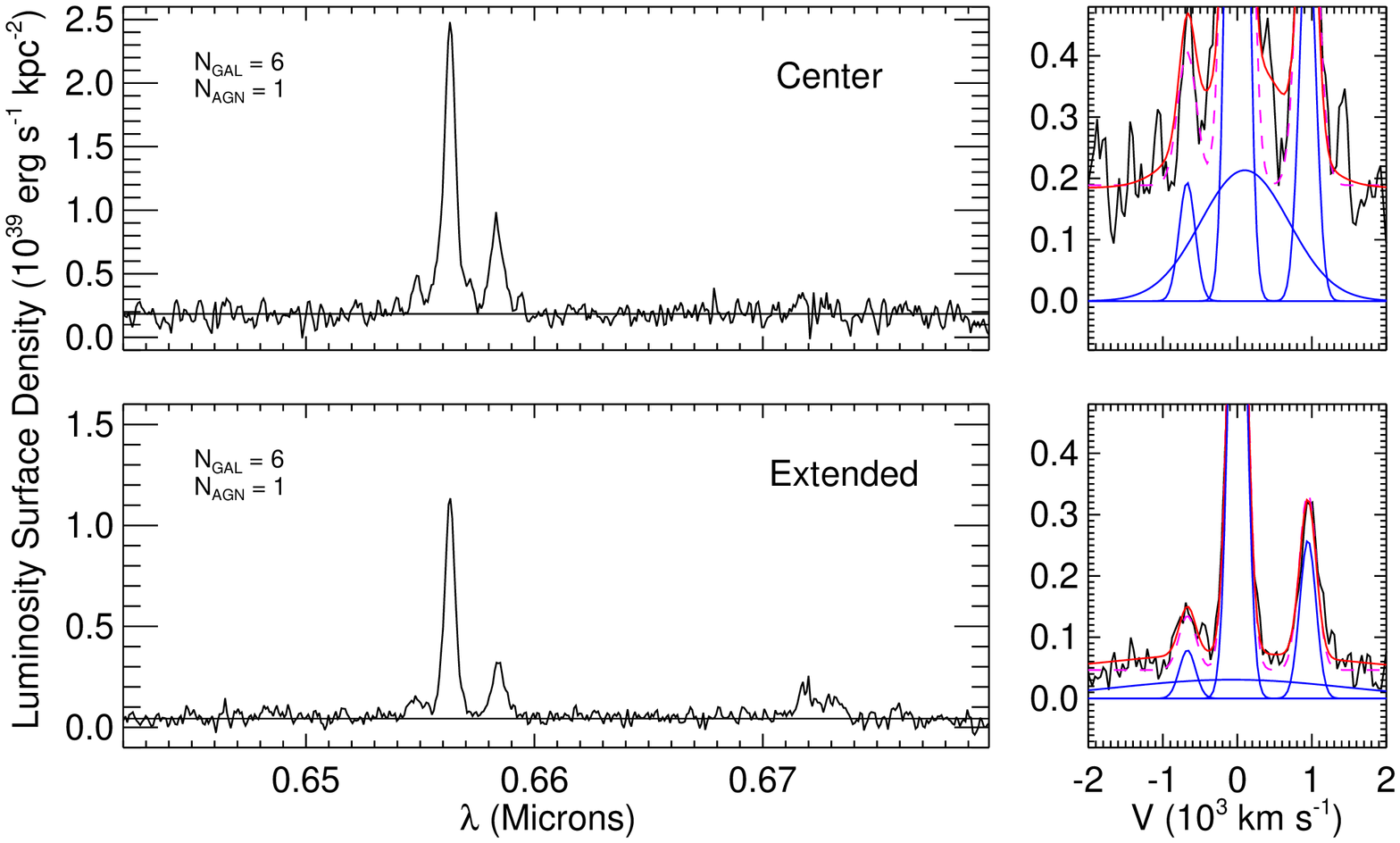}
	\caption{Average spectra, in luminosity surface density, of the central ({\it top panel}) and extended ({\it bottom panel}) regions of well-resolved SINS galaxies.  Panels and colors are as in Figures~\ref{agn}~and~\ref{mstar}.}
	\label{cenext}
\end{figure*}

With the spatially resolved data, we can also compare the integrated spectra from the central ($R~<~3$~kpc) regions of \hz\ galaxies to those from extended ($R~=~3-15$~kpc) regions, in order to determine what regions in these systems are generating broad emission.  For this analysis, we use galaxies from the intermediate and high mass bins of Figure~\ref{mstar} in which the intensity distribution of the stellar continuum defines a clear center of the system (totalling $6$ systems).  The average spectra of the central and extended regions of these systems are shown in Figure~\ref{cenext}, normalized to the spatial area over which the spectra were extracted.

In these spectra, a broad component is preferred by the best-fitting models; however, the significance of this result is low.  The detection of broad emission in galaxy centers at \ztwo\ (with $P_{false} = 8$\%) is somewhat more robust than that in the extended regions ($P_{false} = 14$\%), in which the best fit broad component is very shallow.  If real, the broad feature in the extended regions accounts for a comparable fraction of the total \ha\ luminosity to that in the central regions (Table~\ref{tab:results}).  Tests of simulated galaxies indicate that such a broad line in the extended regions cannot be reproduced by a nuclear point-source of broad emission (i.e. AGN) broadened by the PSF.

\begin{table*}
\caption{Results from Line Fitting}
\label{tab:results}
\begin{center}
\begin{minipage}{18cm}
\begin{center}
\begin{tabular*}{18cm}{llccccccc}
\hline
\ \ \ \ \ Subsample				&\ \ Figure		& $L_{H\alpha,broad}$		& $f_{broad}\ ^a$		& v$_{off}\ ^b$ 	& FWHM$_{H\alpha,broad}$	& $P_{false}$	& $\chi^2_{dof}$ 	& $\chi^2_{dof,narrow}\ ^c$ \\
							& Reference	& ($10^{41}$ erg s$^{-1}$)	& 					& (km s$^{-1}$)			& (km s$^{-1}$)				& 			&				& \\
\hline
\hline
All 							& 1, top 		& $4.0^{+1.1}_{-1.0}$\ \ \	& $0.28^{+0.04}_{-0.08}$ 		& $18^{+79}_{-73}$		& $1632^{+445}_{-301}$		& 0.02	& 1.8		& 4.9 \\	
All, broad \ha\ and [N{\tiny II}] $^d$             & 1, top                & $5.9^{+1.0}_{-1.4}$\ \ \        & $0.42^{+0.08}_{-0.08}$            & $8^{+14}_{-3}$      & $556^{+228}_{-87}$              & 0.01            & 1.4           & 4.9 \\
Non-AGN						& 1, center 	& $3.0^{+0.6}_{-0.4}$\ \ \	& $0.22^{+0.03}_{-0.02}$		& $19^{+82}_{-42}$		& $1329^{+346}_{-139}$		& 0.03	& 1.9		& 3.7 \\
AGN							& 1, bottom  	& $8.2^{+0.7}_{-2.8}$\ \ \	& $0.36^{+0.03}_{-0.09}$		& $42^{+94}_{-29}$		& $2863^{+669}_{-728}$		& 0.007	& 0.9		& 3.0 \\
$M_* < 2 \times 10^{10}\ M_\odot$ 	& 2, top 		& $2.2^{+0.4}_{-0.6}$\ \ \	& $0.21^{+0.03}_{-0.02}$ 		& $26^{+86}_{-30}$ 		& $1051^{+149}_{-169}$		& 0.11	& 0.9		& 1.3  \\
$M_* = 2-7 \times 10^{10}\ M_\odot$ & 2, center 	& $2.9^{+0.6}_{-0.7}$\ \ \	& $0.22^{+0.06}_{-0.02}$ 		& $11^{+71}_{-70}$		& $1425^{+443}_{-238}$		& 0.05	& 1.4		& 2.0  \\
$M_* > 7 \times 10^{10}\ M_\odot$	& 2, bottom	& $6.7^{+2.9}_{-2.3}$\ \ \	& $0.31^{+0.04}_{-0.11}$		& $20^{+82}_{-42}$		& $2183^{+392}_{-758}$		& 0.03	& 1.3		& 3.6 \\
Center						& 3, top 		& $5.1^{+2.8}_{-1.7}\ ^e$	& $0.24^{+0.13}_{-0.04}$		& $9^{+80}_{-27}$		& $1564^{+601}_{-544}$		& 0.08	& 1.3		& 1.9 \\
Extended						& 3, bottom 	& $1.9^{+0.6}_{-0.5}\ ^e$	& $0.22^{+0.06}_{-0.02}$		& $10^{+69}_{-39}$		& $1508^{+725}_{-244}$		& 0.14	& 0.9		& 1.2 \\
\hline
\end{tabular*}
\end{center}
$^a$ Ratio of \ha\ luminosity in broad component to total \ha\ (narrow+broad) luminosity. \\
$^b$ Velocity offset of broad \ha\ feature from narrow \ha. \\
$^c$ Reduced $\chi^2$ for a fit assuming only narrow emission lines. \\
$^d$ Most components of the fit ($v_{off}$, FWHM$_{H\alpha,broad}$), and $f_{broad}$) are constrained in the fit to be identical for the permitted and forbidden lines.  The luminosity ($L_{H\alpha,broad}$) is quoted only for the broad component of the \ha\ line. \\
$^e$ Units are $10^{39}$ erg s$^{-1}$ kpc$^{-2}$.
\end{minipage}
\end{center}
\end{table*}

In the spectra shown in Figure~\ref{cenext}, we note that the average \nii/\ha\ ratio is comparable in the central and extended regions, despite the difference in broad emission component, in support of the above interpretation of the \nii/\ha\ feature as primarily reflecting the metallicity of these systems.  However, this should not be interpreted as a lack of metallicity gradient in these galaxies, since the spectra shown here are averaged over a number of galaxies, in each of which the extended region is spatially integrated over a large range in radii.  Detailed studies of metallicity gradients within and between individual SINS systems will be presented in Buschkamp et al. (in prep).

\section{Discussion}	
\label{Discussion}

The low luminosity broad emission seen in our SINS galaxies can be interpreted in one of two ways: either as evidence of large-scale galactic outflows, presumably driven by the galaxies' very high star formation rates, or as a tracers of the broad-line regions surrounding AGN in these early galaxies.  The current data do not allow us to robustly distinguish between these two scenarios empirically; in the following, we therefore examine both in detail.

\subsection{Broad Emission from Starburst-Driven Winds}

One possible explanation of the broad emission is a starburst-driven wind.   This scenario is in keeping with the positive correlation between broad emission and stellar mass (and therefore star formation rate) seen in Figure~\ref{mstar} and with the possible broad emission from non-nuclear regions seen in Figure~\ref{cenext}.  Moreover, starburst-driven winds are expected to be ubiquitous in the rapidly star-forming populations common at high-redshift \citep[e.g.][]{Pet+01,Shapley+03,Sma+03,Wei+09} and are probably expelling mass from their host galaxies at rates comparable to the star formation rate \citep[below; see also][]{Mar99,Pet+00,Mar03,Erb08,Wei+09}.

At low redshift, star-forming systems are known to drive galactic winds with observable signatures in the wings of the permitted and forbidden emission lines.   In dwarf starburst galaxies, \citet{Wes+07a} find that the broad wings of the emission lines can be modeled as a second Gaussian component with FWHM $\leq$ 300 \kms.  In contrast, observations of the more massive and more rapidly star-forming IR-luminous galaxy population reveal higher FWHM in the broad wings (300$-$800 \kms; \citealt{ArrColCle01}), whose line widths and $f_{broad}$ are comparable to those observed in the SINS galaxies (550 \kms; see Table~\ref{tab:results}; \citealt{Arm+89,Arm+90,LehHec96}).

\begin{figure}
	\centering
	\includegraphics[width=8cm]{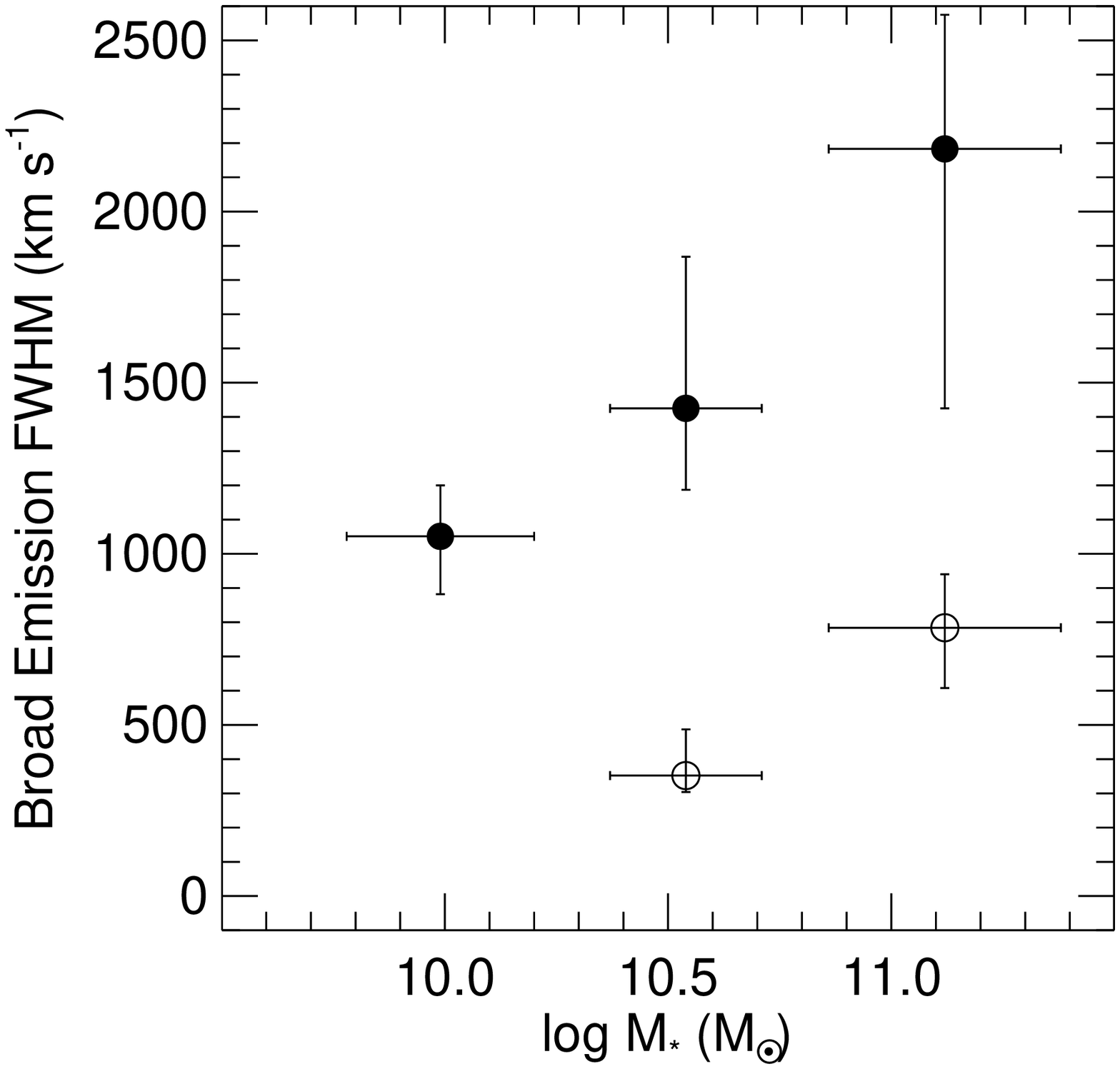}
	\caption{FWHM of the broad component in the SINS galaxies in the three stellar mass bins defined in Figure~\ref{mstar}, as modeled by a single broad \ha\ line ({\it filled circles}) and by broad wings on both the \ha\ and \nii\ lines ({\it open circles}).  In the latter model, the low mass bin is not well constrained due to the low S/N of the \nii\ feature (see Figure~\ref{mstar}), resulting in a fit that strongly prefers a broad line for only the high S/N (\ha) line; we therefore omit this bin from the plot.}
	\label{fwhm}
\end{figure}

Indeed, the relationship between wind speed and SFR (and the equivalent properties, galaxy mass and $B$-band magnitude) has been demonstrated by \citet{RupVeiSan05b}, \citet{Mar05}, and \citet{Tre+07}, respectively.  In particular, \citet{LehHec96} and \citet{RupVeiSan05b} showed that systems with higher SFR have faster winds, whose velocities exceed $\sim 1000$ \kms\ in ionized gas tracers.  We recover similar trends and velocities with the three mass bins of the SINS data, as shown in Figure~\ref{fwhm}.  Note, however, that this Figure should not be directly compared with those in the works listed above, in which the wind speeds are probed with interstellar absorption features, whose velocities are typically lower than those of ionized emission line gas \citep[e.g.][]{Vei+05,RupVeiSan05b}.  Nevertheless, the similarity between the properties of local winds and those of the broad wings seen in the SINS galaxies makes it plausible that this emission is due to the galactic winds that certainly exist in these systems.

These broad wings can be produced via several mechanisms, including turbulent mixing of the hot wind fluid and cool gas clumps at the base of the wind \citep{Wes+07a,Wes+07c} or within the large-scale bubble along the galaxy minor axis \citep[e.g.][]{Arm+90,ArrColCle01}.  In the idealized model of a wind plowing into a spherically symmetric ambient medium, \citet{HecLehArm93} provide analytic expressions for the velocity and \ha\ luminosity of the shocked gas as functions of the bolometric luminosity of the star-formation event, the duration of the event, and the ISM density.  Converting the SFR ($\sim$~100~\msunyr) of the SINS galaxies to a bolometric luminosity and accounting for the correspondingly denser ISM than in local spiral galaxies (by a factor of 10$-$30; \citealt{Bou+07}), these relations predict a \ha\ luminosity of 0.7$-$10~$\times$~10$^{42}$ erg s$^{-1}$ generated from gas moving at $\sim$~70$-$300 \kms.  These properties are broadly consistent with the high-velocity wings observed in the SINS galaxies.

However, our spectra can also be fit by a broad \ha\ line, and we therefore explore whether such a feature could likewise be generated in galactic winds.  For simplicity, we assume that these winds are powered by supernova remnants (SNR), in local examples of which broad \ha\ (FWHM = 500$-$8000 \kms) is observed throughout the Sedov-Taylor expansion phase, the result of charge exchange of the electrons from slow neutral atoms to the fast post-shock protons \citep[e.g.][]{CheRay78,Smi+91,HenSun08}.  Although such broad \ha\ emission is not observed on the galactic scale in local starbursting systems, we note that none of these systems are appropriate analogs to the star formation mode that dominates at \ztwo.  Only local ULIRGs have SFR comparable to that in the SINS galaxies, and the ionized gas emission in ULIRGs suffers significantly more extinction ($A_V =$~5$-$1000; \citealt{Gen+98}) than in SINS galaxies ($A_V \sim$~1).  We therefore briefly examine the possibility that galactic winds at \ztwo\ emit broad \ha\ lines via a superposition of SNR.

A simple test of this scenario is whether there are a sufficient number of SNR in the SINS galaxies to drive the observed broad \ha\ luminosity.  Locally, SNR are observed to have broad \ha\ luminosities of 10$^{30} -$10$^{34}$~erg~s$^{-1}$, generated by change exchange as the shock encounters the ISM and therefore proportional to the square of the gas density.  In the SINS galaxies, the high SFR ($\sim$~100~\msunyr) and dense ISM (10$-$30 times denser than in local spiral galaxies) imply an increase in broad \ha\ luminosity per SNR by a factor of 100$-$1000, to 10$^{32}-$10$^{37}$~erg~s$^{-1}$.  The SFR in the SINS galaxies yields $\sim$~1 supernova explosion per year, each of which will have expected Sedov-Taylor lifetimes of $\sim$~10$^4$ yr in the dense ISM, resulting in roughly 10$^4$ SNR radiating broad \ha\ at any given time in the average SINS galaxy, for a total broad \ha\ luminosity of 10$^{36} - $10$^{41}$ erg s$^{-1}$.  This number is lower than the observed broad \ha\ luminosity in the SINS galaxies by a factor of a few (see Table~\ref{tab:results}); however, real galactic winds penetrate much further through their host galaxies' ISM than do the sum of individual SNR, implying that additional broad \ha\ emission is expected from the interaction of the large-scale winds with the ambient medium.  This effect brings the predictions into even closer agreement with the observations and thus makes this mechanism an energetically feasible explanation for the luminosity and FWHM of the broad emission in the SINS galaxies.

Regardless of whether the broad emission is a broad \ha\ line or broad wings on all emission lines, if the emission is in fact due to starburst-driven winds, it is worth examining what the fate of this high-velocity gas may be.  The average SINS galaxy (top panel of Figure~\ref{agn}) has either broad wings in \ha\ and \nii\ with velocity dispersions of $\sim$ 250 \kms\ or a broad \ha\ line with velocity dispersion $\sim$~700 \kms.  We can compare these values directly with the escape velocity $v_{esc} \approx$~450~\kms\ for a typical SINS galaxy ($\langle {\rm M_{dyn}} \rangle = 8 \times 10^{10}$~\msun, $\langle R_{1/2} \rangle =$~3.4~kpc; \citealt{For+09}).  A significant fraction (7\%\ or 55\%, respectively) of the high-velocity gas has velocity exceeding the host galaxy's escape velocity; assuming that this gas is distributed throughout the star-forming disk, a non-negligible amount of it should be expected to escape the galaxy.  Furthermore, if the surrounding dark matter halo has a flat rotation curve, this gas would also be expected to escape the halo into the intergalactic medium.

We can then estimate the mass outflow rate that would correspond to such superwinds.  The escaping \ha-emitting gas is $\sim$~3$-$17\%\ of the total \ha-emitting gas in the SINS galaxies (i.e. 30\%\ of the emission is broad and 55\%\ of the broad emission escapes).  Assuming that the fraction of gas in the ionized phase is roughly the same in the galaxies' star-forming disks and in the outflows (compare $\sim$~1\%\ in the Milky Way to $\sim$~0.1$-$1\%\ in winds; \citealt{Vei+05}), this implies that $\sim$~3$-$17\%\ of the galaxies' gas reservoirs are being expelled by the observed star-forming event.  The dynamical times associated with these outflows can be approximated as the radius of the star formation event divided by the velocity of the flow; for the SINS galaxies, this yields dynamical times of $\sim$~10~Myr.  The average SINS galaxy has a dynamical mass of 8 $\times$ 10$^{10}$ \msun\ and a gas fraction of 0.2$-$0.4 \citep[e.g.][]{Erb+06b,Bou+07}, yielding an expected outflow rate of 50$-$500 \msunyr.  This is consistent with the results of \citet{Erb08}, who argue that the observed \ztwo\ mass-metallicity relationship requires outflow rates slightly larger than the SFR (SFR~$=$~1$-$800~\msunyr\ with median 72~\msunyr\ in the SINS sample; \citealt{For+09}).  If the broad emission we observe is due to galactic winds at \ztwo, we thus find mass outflow rates consistent with the observed metallicity evolution of these galaxies.

\subsection{Broad Emission from Active SMBHs}

\begin{figure*}
	\centering
	\includegraphics[width=5cm,angle=90]{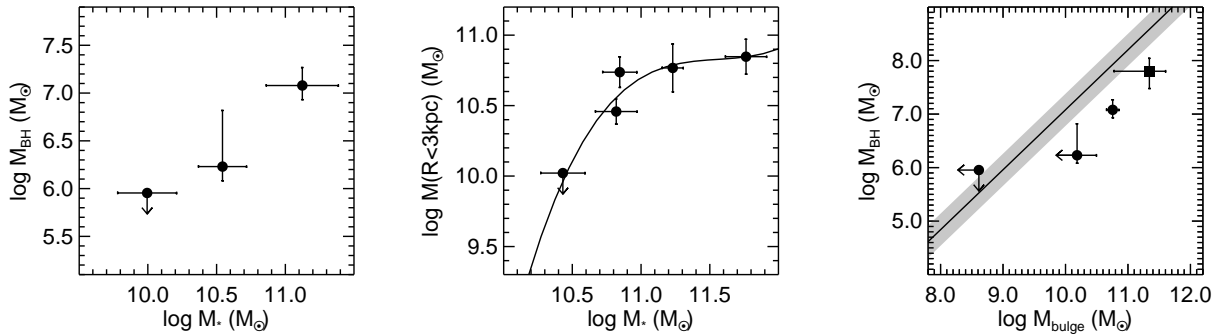}
         \caption{{\it Left}: Estimated black hole masses (derived with virial estimates from the broad \ha\ emission line) against average stellar mass (derived with SED modeling; \citealt{For+09}) in each mass bin.  {\it Center}: Relation between central mass concentration (dynamical mass within $3$ kpc) and stellar mass for the five galaxies modeled in detail by \citet{Gen+08}, along with the best-fitting third-order polynomial.  {\it Right}: Estimated black hole masses against estimated bulge masses ({\it circles;} \mbulge\ estimated from stellar masses shown in left panel and polynomial shown in center panel) for the three mass bins shown in Figure~\ref{mstar}.  Overplotted ({\it square}) is the average BH mass measured in \ztwo\ sub-mm galaxies by \citet{Ale+08}; for comparison with our central mass concentration (dynamical mass within $3$ kpc), we adopt the stellar mass within $\sim 4$ kpc measured by \citet{Ale+08} as the ``bulge" mass, with the error bar extending to include all ``bulge" masses estimated by those authors.  The \mbh-\mbulge\ relation for the local Universe and its scatter are indicated with the solid line and shading \citep{HarRix04}.}
	\label{mbh}
\end{figure*}

Another possible, and common, interpretation of broad emission lines are as signatures of nuclear activity.  For our galaxies, this interpretation is supported by the strength of the broad \ha\ emission increasing with stellar mass (and therefore bulge mass and possibly black hole mass) seen in Figure~\ref{mstar}, by the less luminous broad emission in the non-active systems (the driving mechanism being ``turned off" or obscured; Figures~\ref{agn}~and~\ref{mstar}), and by the more significant detection of broad emission in the centers of galaxies than in their non-nuclear regions.  Moreover, studies of \ztwo\ star-forming galaxies in infrared and X-ray emission suggest that active nuclei may be a common feature of this population \citep{Dad+07b}.

Early work in the local Universe has shown that broad \ha\ is by definition omnipresent in Type 1 AGN, in which the broad-line regions around the nuclei are unobscured.  Moreover, the kinematics and sizes of these regions can be used to infer ``virial" supermassive black hole (SMBH) masses, via calibrated relations between the observed AGN continuum luminosities at 5100 \AA\ ($L_{5100}$) and the sizes of the BLR \citep{Kas+00}, which can then be used in combination with the width of the broad emission lines to estimate SMBH masses \citep{Ves02}.  \citet{GreHo05} have additionally related $L_{5100}$ to the luminosity of the broad emission lines, making it possible to measure virial black hole masses using only a single broad line, 

\begin{eqnarray}
 {\rm M_{BH}} = (2.0^{+0.4}_{-0.3})\ \times\ 10^6 
 \quad \quad \quad \quad \quad \quad \quad \quad \quad \quad \quad \quad \quad \nonumber \\
  \left( \frac{L_{\rm H\alpha}}{10^{42}\ {\rm erg\ s^{-1}}} \right)^{0.55 \pm 0.02} 
  \left( \frac{\rm FWHM_{ H\alpha}}{10^3\ {\rm km\ s^{-1}}} \right)^{2.06 \pm 0.06} 
  {\rm M_\odot} \ . \nonumber \\
  \ 
\label{eq:mbh}
\end{eqnarray}

Such virial black hole mass estimates have been verified against stellar and gas kinematic determinations of \mbh\ and found to be accurate to within a factor of $\sim 3$ \citep[][but see also \citealt{Mar+08}]{Onk+07,HicMal08,Net09}.  As a result, they have been utilized in the \hz\ Universe to probe the masses of SMBHs in quasars \citep[e.g.][]{WilMcLurJar03,McLurDun04,Ves04} and in the sub-mm-bright galaxy population \citep{Ale+08}.

Additionally, the broad-line \ha\ luminosity can also be used to probe the accretion rate of SMBHs.  Assuming the AGN continuum luminosity $L_{5100}$ is roughly $1 / 10$ the AGN bolometric luminosity $L_{\rm bol}$, the accretion rate can be estimated by $\dot{\rm M}_ {\rm BH} = L_{\rm bol} / ( \eta c^2)$ \citep{LaMur+07}.  Using a typical value of $\eta \sim 0.1$ and the calibrated relationship between $L_{\rm H\alpha}$ and $L_{5100}$ \citep{GreHo05}, we can then estimate \mdotbh.  Although approximate, this estimate of \mdotbh\ and the estimate of \mbh\ derived using equation~\ref{eq:mbh} nevertheless allow us to study both the putative black hole masses in our SINS galaxies and their accretion rates using our measurements of the broad \ha\ features.

With this procedure, we derive for the average SINS spectrum (top row in Figure~\ref{agn}) a black hole mass of \mbh~$= 4^{+3}_{-2} \times 10^6$~\msun\ with an Eddington ratio of $\sim 0.2$.  Furthermore, we infer black hole masses of \mbh~$< 9 \times 10^5$~\msun, \mbh~$= 2^{+2}_{-1} \times 10^6$~\msun, and \mbh~$= 1^{+1}_{-0.5} \times 10^7$~\msun\ in each of the three stellar mass bins, respectively.  In the latter two bins, the Eddington ratios are estimated to be 0.3 and 0.4, respectively.

We note that the \ha\ luminosities used in these black hole mass estimates have not been corrected for extinction, and any such correction would increase the derived masses.  However, broad emission lines associated with Type 1-1.5 AGN in local galaxies are consistent with no additional extinction along the line-of-sight to the BLRs \citep[e.g.][]{RheLar00,Alo-Her+03,GreHo05}.  Likewise, at \ztwo, \citet{Ale+08} have used the \ha/\hb\ broad-line Balmer decrement to measure only small amounts of nuclear extinction ($A_V \approx 1.2$) in their dust-rich sub-mm population.  These authors postulate that the plentiful dust in the sub-mm galaxies obscures regions of star-formation and not the BLRs.  In the SINS galaxies, the average extinction in the star-forming regions is $A_{\rm H\alpha} \sim$~0.8; if the BLR is similarly obscured, correcting for this effect would yield an increase of at most a factor of 2 in broad \ha\ luminosity and, from equation~\ref{eq:mbh}, a factor $\sim$~1.5 in derived black hole mass.  This suggests that it is unlikely that the black hole masses measured here suffer significantly from extinction.

It is then naturally interesting to compare the estimated black hole masses with large-scale galaxy properties.  In the left panel of Figure~\ref{mbh}, we find the expected trend that black holes of increasing mass are found in galaxies of increasing stellar mass, with black-hole-to-stellar mass ratios of $\sim$~7~$\times$~10$^{-5}$.  For such black holes to be consistent with local scaling relations between black holes and bulges (\mbh/\mbulge~$\sim$~10$^{-3}$), the bulge-to-total ratio $B/T$ of these systems would need to be quite small ($\lesssim$~0.07).  This is in marked contrast to results from detailed dynamical modeling of the five SINS galaxies with the highest-resolution observations \citep{Gen+08}, which yield $B/T =$~0.15$-$0.4 (center panel of Figure~\ref{mbh}).  Parameterizing these results with a simple polynomial, we estimate ``bulge" masses ($\sim$~dynamical mass within 3~kpc) for the three mass bins, albeit with large uncertainty.  The resulting $B/T$ for these mass bins are $<$~0.1, $<$~0.4, and 0.4, respectively.  These bulge masses are then plotted against our measured black hole masses in the right panel of Figure~\ref{mbh}.

Comparing to the local \mbh-\mbulge\ relation measured by \citet{HarRix04}, we find that our \ztwo\ galaxies lie significantly below the local relation, implying that black holes in the star-forming galaxies in the early Universe may have lagged significantly behind their host bulges in assembly.  \citet{Gen+08} have shown that \ztwo\ is, for many of these systems, an era of bulge formation via smooth but rapid secular processes, during which massive bulges are assembled from large star-forming clumps (M~$\sim 10^8-10^9$~\msun) on timescales of $\lesssim$~1~Gyr.  \citet{ElmBouElm08} have demonstrated in simulations that, assuming each such clump contains a black hole of 10$^{-3}$ of its total mass, these black holes would migrate to the galaxy center with their host clumps and form central SMBHs that are somewhat under-massive for the resulting bulges.  The location of our \ztwo\ systems significantly below the local relation likewise suggests that the bulges in these galaxies form first, through rapid secular processes, with the assembly of the central SMBHs following later.

The timescale for this final SMBH growth can be estimated for the high mass bin, in which the galaxy bulges are probably largely in place at \ztwo\ \citep{Gen+08}.  Accretion onto the black hole at the current rate ($0.4\ \dot{\rm M}_{Edd} \sim 5 \times 10^{-2}$ \msunyr) will bring these galaxies onto the local \mbh-\mbulge\ relation (\mbh~$\sim$ 10$^8$~\msun) in $\sim$~2~Gyr.  Similarly, galaxies in the intermediate mass bin will acquire SMBHs with final masses $\sim$~2~$\times$~10$^7$~\msun\ in $\lesssim$~1~Gyr.  The evolutionary link between these final SMBHs at $z =$~0 (\mbh~$\gtrsim 2 \times 10^7$~\msun) and their putative \ztwo\ host galaxies (\mstar~$> 10^{10}$~\msun) is supported by the similar space densities of these two populations ($\sim 3 \times 10^{-3}\ h_{70}^3$~Mpc$^{-3}$ and $2 \times 10^{-3}\ h_{70}^3$~Mpc$^{-3}$, respectively; \citealt{McLurDun04,Dad+04a,Dad+05,Red+05,Gra+07}).  Moreover, the final black holes (\mbh~$\gtrsim 2 \times 10^7$~\msun) are found at $z =$~0 in elliptical and bulge-dominated spiral galaxies \citep[e.g.][]{Tre+02,Mar+04}, consistent with the probable descendants of the \ztwo\ star-forming galaxy population \citep{Genel+08,Con+08}.   This evidence thus confirm the plausibility of the rapid bulge formation seen at \ztwo\ in the SINS galaxies \citep{Gen+08} being followed by a few Gyr of rapid SMBH assembly, ultimately resulting in spheroids and bulges that obey the local \mbh-\mbulge\ relation.

\citet{Ale+08} found similar delayed SMBH formation in the sub-mm galaxy population at \ztwo\ (see Figure~\ref{mbh}), suggesting that the time lag between black holes and bulges may be a common phenomenon in rapidly forming galaxies at high redshift.  However, this trend is not universal; quasars and radio galaxies at similar redshifts are suspected to lie {\it above} local black hole scaling relations, with black holes that are over-massive for their host bulges by up to and exceeding an order of magnitude \citep[e.g.][but see also \citealt{Shi+03}]{Wal+04,Shi+06,McLur+06,Pen+06,Mai+07}.  These systems populate the highest mass end of the black hole mass function, with black hole masses of $> 10^8-10^9$ \msun\ already in place at \ztwo.  It may therefore be that black holes in these different mass/activity regimes grow in very different circumstances and consequently relate to their bulges very differently.  If this is the case, the challenge is then to locate the mechanism(s) that bring these varied high-redshift formation processes together into the black hole scaling relations observed at $z = 0$.

\section{Conclusions}
\label{Conclu}

In stacked, average spectra of SINS \ztwo\ star-forming galaxies, we have detected broad emission underneath the much brighter narrow \ha\ and \nii\ emission lines.  This broad emission accounts for $\sim$ 30\%\ of the total \ha\ luminosity of these galaxies and can be parameterized equally well with a single broad \ha\ line (``broad line" of FWHM $\sim$ 1500 \kms) and with a two Gaussian fit to both the permitted and forbidden lines (``broad wings" of FWHM $\sim$ 550 \kms).  The luminosity and FWHM of the broad component increases with increasing galaxy mass and therefore with SFR.  This broad component is found both in known AGN and in stacked spectra of systems that have not been previously identified as AGN.  There is some evidence that the broad emission is more luminous in galaxy centers, as opposed to in the outer regions, but the significance of these detections are low.

We cannot empirically determine whether this broad emission is due to high-velocity galactic winds and the associated shocks or to the BLR emission of AGN.  In the former case, we find that simple scaling arguments show that the luminosity and FWHM of the broad emission can plausibly be accounted for via shocking of the ambient interstellar media from supernovae-driven galactic winds.  These winds would then be ejecting matter from the host galaxy at rates slightly exceeding the star formation rate, in keeping with expectations from the metallicity evolution of these galaxies and with ultraviolet interstellar absorption-line studies at similar redshifts.

On the other hand, the broad emission may be generated in a BLR; in this case, we can estimate the black hole masses and luminosities necessary to fuel the observed emission for each of three galaxy mass bins.  We find that the measured SMBH masses correlate with the host galaxy masses, as expected from local scaling relations, but that the SMBHs are significantly under-massive for their bulges when compared with local relations.  While this result has large uncertainties, it is consistent with the emerging picture of galaxy assembly at \ztwo, in which a gas-rich disk fragments into large ($\ge 1$ kpc) super-star-forming clumps that then migrate into the galaxy center on Gyr-timescales to form a nascent bulge.  The bulge would then form first through this process and only later completely assemble its black hole.

The obvious direction for future research is to determine the source of the broad \ha\ emission in high-redshift star-forming galaxies.  This will most likely require detailed examination of individual galaxies.  Among the diagnostics that will be useful for this task are comparisons with X-ray data and deep integrations in rest-frame UV/optical wavebands to spatially resolve e.g. UV interstellar absorption lines and broad Balmer emission.

\section*{Acknowledgments}

We thank the ESO staff, especially those at Paranal Observatory, for their helpful and enthusiastic support during the many observing runs and several years over which the SINS project was carried out.  We also acknowledge the SINFONI and PARSEC teams, whose hard work on the instrument and laser paved the way for the success of the SINS observations.  This paper has additionally benefited significantly from many enlightening conversations with colleagues, including Fr\'ed\'eric Bournaud, Mohan Ganeshalingam, Kevin Heng, Phil Hopkins, Chris McKee, Jeffrey Silverman, and Thea Steele.  Finally, we thank the referee, whose detailed and insightful comments greatly improved the quality of this paper.



 \end{document}